\documentclass[aps,pre,showpacs,superscriptaddress,manuscript]{revtex4-2}
\usepackage{amssymb}
\usepackage{amsthm}
\usepackage{graphicx,subfigure}
\usepackage{epsfig}
\usepackage{multirow}
\usepackage{color}
\usepackage{ulem}

\begin{document}
	
\title{Effect of an electromagnetic field on the barrier crossing rate}

\author{L.~R.~Rahul~Biswas}%
\affiliation{Department of Chemistry,\
	Visva-Bharati, Santiniketan,\
	India, 731235 }%

\author{Shrabani~Mondal}
 \affiliation{Department of Chemistry,\
		Physical Chemistry Section,\
		Jadavpur University, Kolkata,\
            India, 700032}
	
\author{Bidhan~Chandra~Bag}%
 \email{bidhanchandra.bag@visva-bharati.ac.in}
\affiliation{Department of Chemistry,\
	Visva-Bharati, Santiniketan,\
	India, 731235 }



\begin{abstract}
We investigate the spectrum for the rate constant of an electric field-driven charged Brownian particle in the presence of a magnetic field (MF). For the cross fields with low or high values of the cyclotron frequency, an asymmetric splitting of the spectrum occurs with two peaks. Anharmonicity-induced additional splitting may appear around the lower resonating frequency at the intermediate strength of the applied MF. Another observation is that if the magnetic field is tilted from the $z$-direction,  an additional peak appears between the two peaks. The position of the middle peak may be independent of the strength of the applied MF. In some cases, only one peak appears even in the presence of a magnetic field. We explain these observations considering the dynamics around the stable fixed point and determine the position of the peak in the spectrum for the rate constant as a function of the strength of the applied magnetic field. 
Thus the present study may find applications for tuning the conductivity of a solid electrolyte, which is very important in recent technology. Other applications may be in areas such as electromagnetic field-induced modulation of (a) thermally activated tunneling ionization, (b) thermally stimulated ionization, etc.

\end{abstract}


\maketitle

\newpage

{\bf Tuning of conductivity of solid electrolytes by a suitable physical method instead of varying chemical composition of these seems to be an interesting issue in recent technology. The present paper includes a related study. Although the time independent magnetic filed does not work but it can modulate the frequency of a dynamical system. Then we determine the condition at which time dependent electric filed driven barrier crossing rate is maximum in the presence of a constant magnetic field. At an intermediate value of the cyclotron frequency, the nonlinearity of the dynamical system may take an important role in the process of tuning of barrier rate. Symmetries of the potential energy field and the magnetic field also may be important in this process through the interference due to the velocity dependent coupling.}

\section{Introduction}

The study of the properties of solid electrolytes is a crucial issue in recent technology.
The materials have potential applications in a diverse range of all-solid-state devices, such as rechargeable lithium batteries, flexible electrochromic displays, and smart windows \cite{scros,bruce,gray,tarascon,glashan,barnes}. The properties of the electrolytes are tuned by varying chemical composition to a large extent and hence are adapted to specific needs \cite{angel,lilley}. High ionic conductivity is needed for optimizing the glassy electrolytes in various applications. Thus tuning of the conductivity of solid electrolytes seems to be a challenging issue. Then searching for a suitable physical method would be an interesting one. 
In this context, the Lorentz force may play an important role. Although a time-independent magnetic field can not activate a Brownian particle to cross the energy barrier, it may modulate the frequency of the dynamical system. Making use of this notion, the barrier crossing rate of a charged Brownian particle in the presence of a time-independent magnetic field was reported in Refs. \ cite {baura,baura1,filliger}. \textcolor{blue}{Very recently, the escape dynamics of a charged
Brownian particle from a two-dimensional truncated harmonic potential under the influence of
Lorentz force due to an external magnetic field has been reported in Ref. \ cite {abdoli1}.}
For a time-dependent field, the induced electric field may activate a Brownian particle to cross the energy barrier \cite{baura2,mondal,mondal1}. Effects of an MF on hot electron transport in quantum wires have been studied in Ref. \ cite {telang}.
In the recent past, the effect of a time-independent magnetic field on the tunneling process was investigated in both experimental\cite{vdovin} and theoretical studies\cite{perel,amemiya}, respectively. The probability of electron tunneling from a bound to a free state under an
alternating electric field in the presence of a constant magnetic field was calculated in Ref. \ cite {moskalenkoa}. Another relevant study\cite{moskalenkoa1} in this direction is the effect of a magnetic field on thermally stimulated ionization of impurity centers in semiconductors by
submillimeter radiation. Recently, in Ref. \ cite{aquino}, a fluctuating electric field-assisted barrier crossing dynamics was studied at the quasi-deterministic limit. 
Thus, the investigation of barrier crossing dynamics in the presence of both electric and magnetic fields is a worthy issue. The objective of the present study is to calculate the barrier crossing rate of a Brownian particle driven by a time-dependent periodic electric field in the presence of a static MF. We considered a model system with a double-well potential energy field in this context. Calculating the mean lifetime in the left well, we determine a spectrum for the rate constant in the presence of cross fields. If the cyclotron frequency is low or high, then an asymmetric splitting of the spectrum occurs with two peaks. 
Anharmonicity-induced additional splitting appears around the lower resonating frequency at the intermediate strength of the applied magnetic field. Another observation is that if MF is tilted from the $z$-direction, an additional peak appears between the two peaks. The position of the middle peak may not depend on the strength of the applied MF. For the dynamics around the bottom of the well, like the Lorentz force-driven isotropic harmonic oscillator, the position of the middle peak is independent of the strength of the field, whose all components are the same. In some cases, only one peak appears even in the presence of a magnetic field as a signature of interference among the components of motion.

The outlay of the paper is as follows. In Sec. II, we calculate the rate constant in the presence of cross fields. The effect of interference among the components of an applied magnetic field (through the velocity-dependent coupling) on the barrier crossing dynamics is addressed in the next section. The paper is concluded in Sec. IV.

\section{Barrier crossing dynamics in the presence of cross fields: Asymmetric splitting of the spectrum for the rate constant}

To study the effect of electromagnetic field on the barrier crossing dynamics, we start with the cross fields, i.e.,  magnetic field (${\bf B} = (0,0, B )$) is applied along the $z$-direction in the presence of an electric field which is perpendicular to it. Then the relevant equations of motion in the SI unit are \cite{mondal1,baura3}

\begin{equation}\label{eq1}
\dot{x} = u_x   \; \; \;,
\end{equation}

\begin{equation}\label{eq2}
\dot{y} = u_y   \; \; \;,
\end{equation}

\begin{equation}\label{eq2a}
\dot{z} = u_z   \; \; \;,
\end{equation}

\begin{equation}\label{eq3}
\dot{u_x} = -\frac{\partial V}{\partial x} + \Omega u_y + qE_x (t) - \gamma u_x + f_x(t)   \; \; \;,
\end{equation}

\begin{equation}\label{eq4}
\dot{u_y} = -\frac{\partial V}{\partial y} - \Omega u_x + qE_y (t) - \gamma u_y + f_y(t)   \; \; \;,
\end{equation}

\noindent
and

\begin{equation}\label{eq4a}
\dot{u_z} = -\frac{\partial V}{\partial z} - \gamma u_z +f_z
\end{equation}

\noindent
where

\begin{equation}\label{eq5}
V(x,y,z) = a x^4 - b x^2 +  (\omega_y^2 y^2 +\omega_z^2 z^2)/2   \; \; \;,
\end{equation}

\noindent
and

\begin{equation}\label{eq6}
\Omega = \frac{q B}{m} \; \; \;.
\end{equation}

\noindent
In Eq.(\ref{eq5}), the parameters $a$ and $b$ measure the curvature of the potential energy field, the location of the fixed points, and the barrier height, respectively. The remaining parameters $\omega_y$ and $\omega_z$  associated with the potential energy do not affect the location of fixed points and the barrier height; however, the curvature of the potential energy depends on these parameters. \textcolor{blue}{Now we would mention the choice of the potential energy field (\ref{eq5}). The key objective of the present study is to explore the electric field-driven resonant activation in the presence of a magnetic field. Then, one may consider the  escape of a Brownian particle from a well with a finite energy barrier. The top of the energy barrier may correspond to an unstable fixed point in the potential energy field. Thus the field is a non linear one. It is to be noted here that in general, a non linear potential energy field was considered for theoretical investigation on the barrier crossing dynamics\cite{hang}. The investigation finds application in biology, chemistry, and physics. Keeping it in mind, we consider a simple as well as relevant non linear potential energy field. To include the effect of a magnetic field on the barrier crossing dynamics, one must use a multidimensional dynamical system. Then we choose that the escaping Brownian particle is bound harmonically along the perpendicular directions (y-z plane) of barrier crossing. Thus we assume that the relevant macroscopic object is effectively one-dimensional, aligned along the $x$-axis. It may correspond to the hopping of a charged particle between two consecutive wells with a finite energy barrier.  
A relevant well may be a result of electrostatic interaction of the particle with ions (having opposite charge) surrounding it. Thu,s the use of the potential energy function (\ref{eq5}) may give essential features of hopping dynamics of a charged particle in a one-dimensional wire. Additional discussion regarding the use of the function is given in Sec.IIB. Here one may find that it may capture essential features of the escaping of a particle from a well in any direction in the $x-y$ plane. Another important advantage
of the choice of the potential energy field (\ref{eq5}) is that it leads to the study on the barrier crossing dynamics numerically based on the calculation of the mean first passage time. It is to be noted here that based on the determination of the mean escape time 
one may study the escape problem in the presence of a magnetic field using a truncated harmonic potential energy field \cite{abdoli1}.
But it is difficult to define the mean first passage time for this field.} 

We now consider the parameter, $\Omega$, in Eq.(\ref{eq6}). It corresponds to the cyclotron frequency of a particle (with charge $q$ and mass $m$) in the presence of a magnetic field, ${\bf B} = (0, 0, B)$.
The remaining part of the cross fields is represented by  ${\bf E} = (E_x(t), E_y(t),0)$. Finally, we consider the terms in the equations of motion related to the thermal bath. $\gamma$ measures the damping strength. $f_x$ and $f_y$ are the components of the random force. We assume that these are independent white Gaussian noises with zero mean, $\langle f_x(t) \rangle = \langle f_y(t) \rangle=\langle f_z(t) \rangle=0$ and related to the damping strength by the fluctuation-dissipation relation,

\begin{equation}\label{eq7}
\langle f_x(t) f_x(t') \rangle = \langle f_y(t) f_y(t^\prime) \rangle = \langle f_z(t) f_z(t^\prime) \rangle = 2 \gamma k_B T \delta(t - t^\prime)   \; \; \;.
\end{equation}

\noindent
Here $k_B$ is the Boltzmann's constant, and $T$ is the temperature of the thermal bath. 

We are now in a position to demonstrate the spectrum for the rate constant. We choose the components of the time-dependent electric field as

\begin{equation}\label{eq8} 
E_x(t) = E_{0x} \cos (\omega_E t)
\end{equation}

\noindent
and

\begin{equation}\label{eq9}
E_y(t) = E_{0y} \cos (\omega_E t) \; \; \;,
\end{equation}

\noindent
respectively. $E_{0x}$ and  $E_{0y}$ are relevant amplitudes of the respective driving components. Making use of the above relations in Eqs.(\ref{eq3}-\ref{eq4}) we have

\begin{equation}\label{eq10}
\dot{u_x} = - 4 a x^3 + 2 b x^2 - \gamma u_x + \Omega u_y + f_x(t) + q E_{0x} \cos (\omega_E t)  \; \; \;,
\end{equation}

\noindent
and

\begin{equation}\label{eq11}
\dot{u_y} = - \omega_y^2 y - \gamma u_y - \Omega u_x + f_y(t) + q E_{0y} \cos (\omega_E t)   \; \; \;,
\end{equation}

\begin{figure}[!htb]
\includegraphics[width=17cm,angle=0,clip]{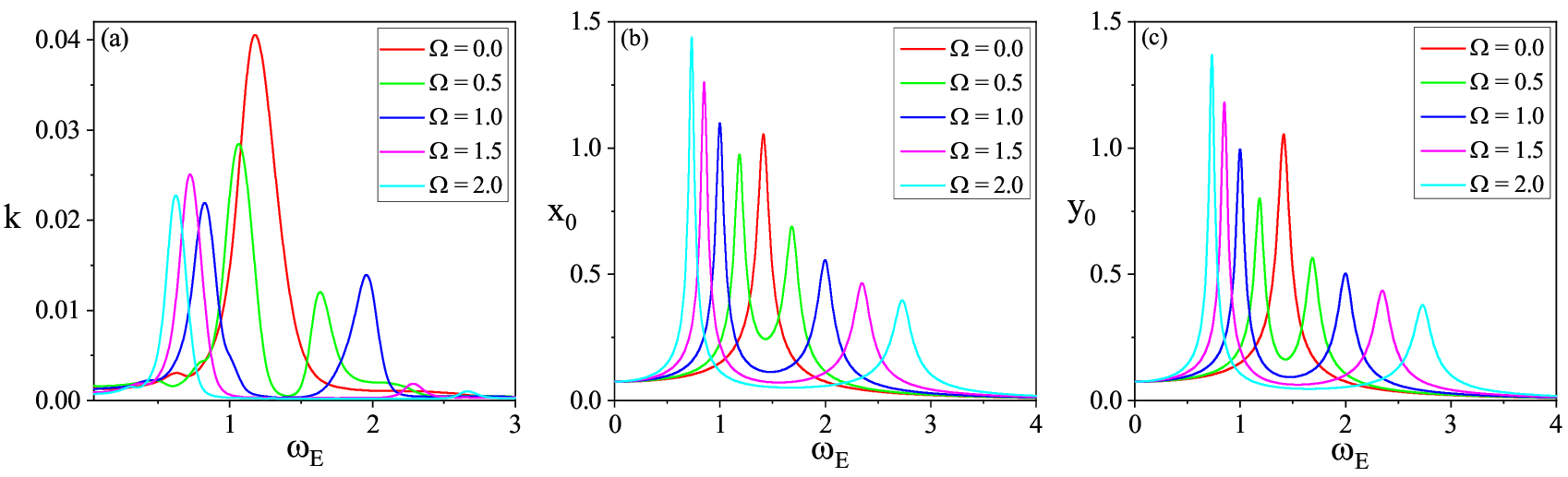}
\caption{Plot of (a) $k$ vs. $\omega_E$ at different values of $\Omega$ for $a = 0.25$, $b = 0.5$, $\omega_y^2 = 2.0$ and $K_BT=0.04$ (b), (c) $x_0$ and $y_0$ vs. $\omega_E$ for different values of $\Omega$ along with the relevant parameter set, $\omega_0^2 = \omega_y^2 = 2.0$. Common parameters are $\gamma = 0.1$ and $E_{0x} = E_{0y} = 0.15$. (Units are arbitrary).}
\label{fig.1}
\end{figure}

\noindent 
Using Heun's method \cite{tor}, we solve these equations along with Eqs. (\ref{eq1}-\ref{eq2}) numerically and calculate the first passage time $t_f$,i.e., the time required for a trajectory that starts from the coordinate ($x = - \sqrt{b/2a}, y = 0$) corresponding to the left minimum (most probable state) of the two-dimensional potential to reach the top of the energy barrier($x = 0, y = 0$) for the first time. For further details, we refer to Ref. \cite{mondal1} where the numerical calculation procedure was discussed in depth. The reliability of the method was well justified in Ref. \cite{mondal}. However, the barrier crossing is assisted by the noise, and therefore, $t_f$ is a statistical quantity. We determine its mean value, $\langle t_f \rangle$, over many realizations of the trajectories such that the mean becomes independent of the number of realizations. In general, we consider 20000 to 25000 trajectories for the statistical averaging. The inverse of the mean first passage time gives the barrier crossing rate constant, $k = \frac{1}{\langle t_f \rangle}$. We demonstrate the variation of the rate constant with  $\omega_E$ in Fig. \ref{fig.1}. In the following subsections, we will consider in detail about the spectrum in this figure.

\subsection{Asymmetric splitting of the spectrum with respect to the position of peaks}

\textcolor{blue}{One may notice a number of features of Fig.1(a). First, the splitting of the rate constant as well as non monotonic probability of barrier crossing in the presence of the magnetic field requires establishing that the relevant dynamical system possesses two natural frequencies.
Since the energy activated process at the bottom is key for the barrier crossing then we consider the dynamics in this regime of the left well. The relevant equations of motion are}

\begin{equation}\label{eq10}
\dot{u_x} = - \omega_0^2x - \gamma u_x + \Omega u_y + f_x(t) + q E_{0x} \cos (\omega_E t)  \; \; \;,
\end{equation}

\noindent
and

\begin{equation}\label{eq11}
\dot{u_y} = - \omega_y^2 y - \gamma u_y - \Omega u_x + f_y(t) + qE_{0y} \cos (\omega_E t)   \; \; \;.
\end{equation}

\noindent
The parameter, $\omega_0$ in Eq.(\ref{eq10}), is the angular frequency for the harmonic motion (along $x$-direction) around the bottom of the well. Since the barrier crossing dynamics for the parameter set corresponding to the peak of the spectrum is mainly controlled by the resonance dynamics, we consider the above equations of motion at the limit $f_x(t) = f_y(t) \simeq 0$,

\begin{equation}\label{eq10a}
\dot{u_x} = - \omega^2x - \gamma u_x + \Omega u_y  + q E_{0y} \cos (\omega_E t)  \; \; \;,
\end{equation}

\noindent
and

\begin{equation}\label{eq11a}
\dot{u_y} = - \omega^2 y - \gamma u_y - \Omega u_x  + q E_{0y} \cos (\omega_E t)   \; \; \;.
\end{equation}

\noindent
Here, we have used $\omega_0 = \omega_y = \omega$. In the next section, we will consider the case with $\omega_0 \neq \omega_y$.
\textcolor{blue}{At low damping limit, solution of the above equations of motion in the absence of driving force can be read as \cite{rahul}}
\textcolor{blue}{
\begin{eqnarray}
x(t) & \simeq & \frac{A_{x0}}{4}p_{0}e^{-\frac{\gamma t}{2}}\left(e^{\frac{Bt}{2\sqrt{A}}}+e^{-\frac{Bt}{2\sqrt{A}}}\right)-\frac{A_{x0}}{4}q_{0}e^{-\frac{\gamma t}{2}}\left(e^{\frac{Bt}{2\sqrt{A}}}-e^{-\frac{Bt}{2\sqrt{A}}}\right)\nonumber \\
\nonumber \\
 & + & \frac{A_{y0}}{4}s_{0}e^{-\frac{\gamma t}{2}}\left(e^{\frac{Bt}{2\sqrt{A}}}-e^{-\frac{Bt}{2\sqrt{A}}}\right)-\frac{A_{y0}}{4}r_{0}e^{-\frac{\gamma t}{2}}\left(e^{\frac{Bt}{2\sqrt{A}}}-e^{-\frac{Bt}{2\sqrt{A}}}\right)\;\;\;\label{eq10b}
\end{eqnarray}}

\noindent and

\textcolor{blue}{\begin{eqnarray}
y(t) & \simeq & \frac{A_{y0}}{4}p_{0}e^{-\frac{\gamma t}{2}}\left(e^{\frac{Bt}{2\sqrt{A}}}+e^{-\frac{Bt}{2\sqrt{A}}}\right)-\frac{A_{y0}}{4}q_{0}e^{-\frac{\gamma t}{2}}\left(e^{\frac{Bt}{2\sqrt{A}}}-e^{-\frac{Bt}{2\sqrt{A}}}\right)\nonumber \\
\nonumber \\
 & + & \frac{A_{x0}}{4}s_{0}e^{-\frac{\gamma t}{2}}\left(e^{\frac{Bt}{2\sqrt{A}}}-e^{-\frac{Bt}{2\sqrt{A}}}\right)-\frac{A_{x0}}{4}r_{0}e^{-\frac{\gamma t}{2}}\left(e^{\frac{Bt}{2\sqrt{A}}}-e^{-\frac{Bt}{2\sqrt{A}}}\right)\;\;\;\label{eq11b}
\end{eqnarray}}

\noindent \textcolor{blue}{where $A_{x0}$ and $A_{y0}$ correspond to initial position of the oscillator,
$A=\omega^{2}-\gamma^{2}/4+\Omega^{2}/4$, $B=-\frac{\gamma\Omega}{2}$, $p_{0}=\cos\Big\{\Big(\sqrt{A}+\frac{\Omega}{2}\Big)t\Big\}+\cos\Big\{\Big(\sqrt{A}-\frac{\Omega}{2}\Big)t\Big\}$,
$q_{0}=\cos\Big\{\Big(\sqrt{A}+\frac{\Omega}{2}\Big)t\Big\}-\cos\Big\{\Big(\sqrt{A}-\frac{\Omega}{2}\Big)t\Big\}$,
$r_{0}=\sin\Big\{\Big(\sqrt{A}+\frac{\Omega}{2}\Big)t\Big\}+\sin\Big\{\Big(\sqrt{A}-\frac{\Omega}{2}\Big)t\Big\}$
and $s_{0}=\sin\Big\{\Big(\sqrt{A}+\frac{\Omega}{2}\Big)t\Big\}-\sin\Big\{\Big(\sqrt{A}-\frac{\Omega}{2}\Big)t\Big\}$.
The above equations imply that the damped oscillation is composed
of two frequencies, $\sqrt{A}+\frac{\Omega}{2}$ and $\sqrt{A}-\frac{\Omega}{2}$,
respectively. Generation of two frequencies is asymmetric in nature with respect to the frequency of the damped harmonic oscillator, $\sqrt{\omega^{2}-\gamma^{2}/4}$. Qualitatively, it may be apparent that the magnetic force induced cyclic motion may favor or oppose depending upon its phase relationship with the cyclic motion in the $x-y$ plane due to the harmonic force field. Then it is expected to appear two peaks will appear at these driving frequencies as a signature of dynamical resonance.
In other words, the probability of barrier crossing at other frequencies around the resonating one is comparatively lower. This is the reason
for the splitting of the rate constant as well as non non-monotonic probability of barrier crossing in the presence of the magnetic field.}

For further discussion, we consider the solution of the equations (\ref{eq10}-\ref{eq11}) of motion for the driven system at the steady state and it is given by 
\cite{rahul}

\begin{equation}\label{eqr3}
x(t) = x_0 \cos(\omega_E t - \phi_1)   \;\;\;,
\end{equation}

\noindent
and

\begin{equation}\label{eqr4}
y(t) = y_0 \cos(\omega_E t - \phi_2)   \;\;\;, 
\end{equation}

\noindent
where

\begin{equation}\label{eqr11}
x_0 = \frac{\sqrt{H_1^2 + H_2^2}}{H_0}   \;\;\;,
\end{equation}

\begin{equation}\label{eqr12}
y_0 = \frac{\sqrt{H_3^2 + H_4^2}}{H_0}   \;\;\;,
\end{equation}

\begin{equation}\label{eqr13}
\tan \phi_1 = \frac{H_2}{H_1}  
\end{equation}

\noindent
and

\begin{equation}\label{eqr14}
\tan \phi_2 = \frac{H_4}{H_3}   \;\;\;
\end{equation}

\noindent
with

\begin{equation}\label{eqr15}
H_0 = \left\{\left(\omega^2 - \omega_E^2\right)^2 - \left(\Omega^2 - \gamma^2\right) \omega_E^2\right\}^2 + 4 \gamma^2 \Omega^2 \omega_E^4   \;\;\;,
\end{equation}

\begin{equation}\label{eqr16}
H_1 = \frac{q}{m} \left(\omega^2 - \omega_E^2\right) \left[\left\{\left(\omega^2 - \omega_E^2\right)^2 - \left(\Omega^2 - \gamma^2\right) \omega_E^2\right\} E_{0x} + 2 \gamma \Omega \omega_E^2 E_{0y} \right]   \;\;\;,
\end{equation}

\begin{equation}\label{eqr17}
H_2 = \frac{q}{m} \omega_E \left[\left\{\left(\omega^2 - \omega_E^2\right)^2 - \left(\Omega^2 - \gamma^2\right) \omega_E^2\right\} \left(\gamma E_{0x} - \Omega E_{0y}\right) + 2 \gamma \Omega \omega_E^2 \left(\gamma E_{0y} + \Omega E_{0x}\right)\right]   \;\;\;,
\end{equation}

\begin{equation}\label{eqr18}
H_3 = \frac{q}{m} \left(\omega^2 - \omega_E^2\right) \left[\left\{\left(\omega^2 - \omega_E^2\right)^2 - \left(\Omega^2 - \gamma^2\right) \omega_E^2\right\} E_{0y} - 2 \gamma \Omega \omega_E^2 E_{0x} \right]   \;\;\;,
\end{equation}

\noindent
and

\begin{equation}\label{eqr19}
H_4 = \frac{q}{m} \omega_E \left[\left\{\left(\omega^2 - \omega_E^2\right)^2 - \left(\Omega^2 - \gamma^2\right) \omega_E^2\right\} \left(\gamma E_{0y} + \Omega E_{0x}\right) - 2 \gamma \Omega \omega_E^2 \left(\gamma E_{0x} - \Omega E_{0y}\right)\right]   \;\;\;.
\end{equation}

\noindent
Variation of $x_0$ and $y_0$ with driving frequency corresponding to Fig.~\ref{fig.1}(a) is demonstrated in panels (b) and (c) of the same figure. It seems that there is a beautiful correspondence between panels (a) and (b) as well (c) regarding locations of maxima. Thus the peaks in the panel (a) are due to the electric field-driven resonant activation. 
Following Ref.~\cite{rahul}, we determine the positions of maxima in this panel at the limit $\gamma\rightarrow 0$ as

\begin{equation}\label{eqr31}
\omega_L \simeq \sqrt{\omega^2 + \frac{\Omega^2}{4}} - \frac{\Omega}{2}   \;\;\;,
\end{equation}

\noindent
and

\begin{equation}\label{eqr32}
\omega_R \simeq \sqrt{\omega^2 + \frac{\Omega^2}{4}} + \frac{\Omega}{2}   \;\;\;.
\end{equation}

\noindent
Here $\omega_L$ and  $\omega_R$ correspond to the driving frequencies for the left and the right peaks, respectively. \textcolor{blue}{These are corresponding to $\sqrt{A}-\frac{\Omega}{2}$ and $\sqrt{A}+\frac{\Omega}{2}$, respectively as suggested by the solutions (\ref{eq10b}-\ref{eq11b})). Thus the  
magnetic field induced asymmetric splitting of the spectrum with respect to the position of peaks is suggested by dynamics around the bottom of the potential energy well.} In Table \ref{tab.1}, the theoretically calculated peak position is compared with the numerical result. It shows a fair agreement between these for $\omega_R$. For 
$\omega_L$, the theoretically calculated value is a little bit higher compared to the exact one. It implies that the non-linearity is significant in decreasing the value of $\omega_L$. 

\begin{table}[ht]
\caption{Comparison between theoretically calculated peak position and the exact result}
\begin{center}
\begin{tabular}{|c|c|c|c|c|}
\hline
Value of &
\multicolumn{2}{|c|}{Resonance at $\omega_L$} &
\multicolumn{2}{|c|}{Resonance at $\omega_R$} \\
\cline{2-5}
$\Omega$ & Theoretical & Exact & Theoretical & Exact \\ 
\hline
0.5 & 1.190 & 1.070 & 1.680 & 1.650 \\
\hline
1.0 & 1.002 & 0.860 & 1.997 & 1.950 \\
\hline
1.5 & 0.852 & 0.729 & 2.348 & 2.300 \\
\hline
2.0 & 0.732 & 0.629 & 2.730 & 2.680 \\
\hline
\end{tabular}
\end{center}
\label{tab.1}
\end{table}

\subsubsection*{Additional splitting of the spectrum around the lower resonating frequency: Signature of non-linearity of the potential energy field}

\begin{figure}[!htb]
\includegraphics[width=16.5cm,angle=0,clip]{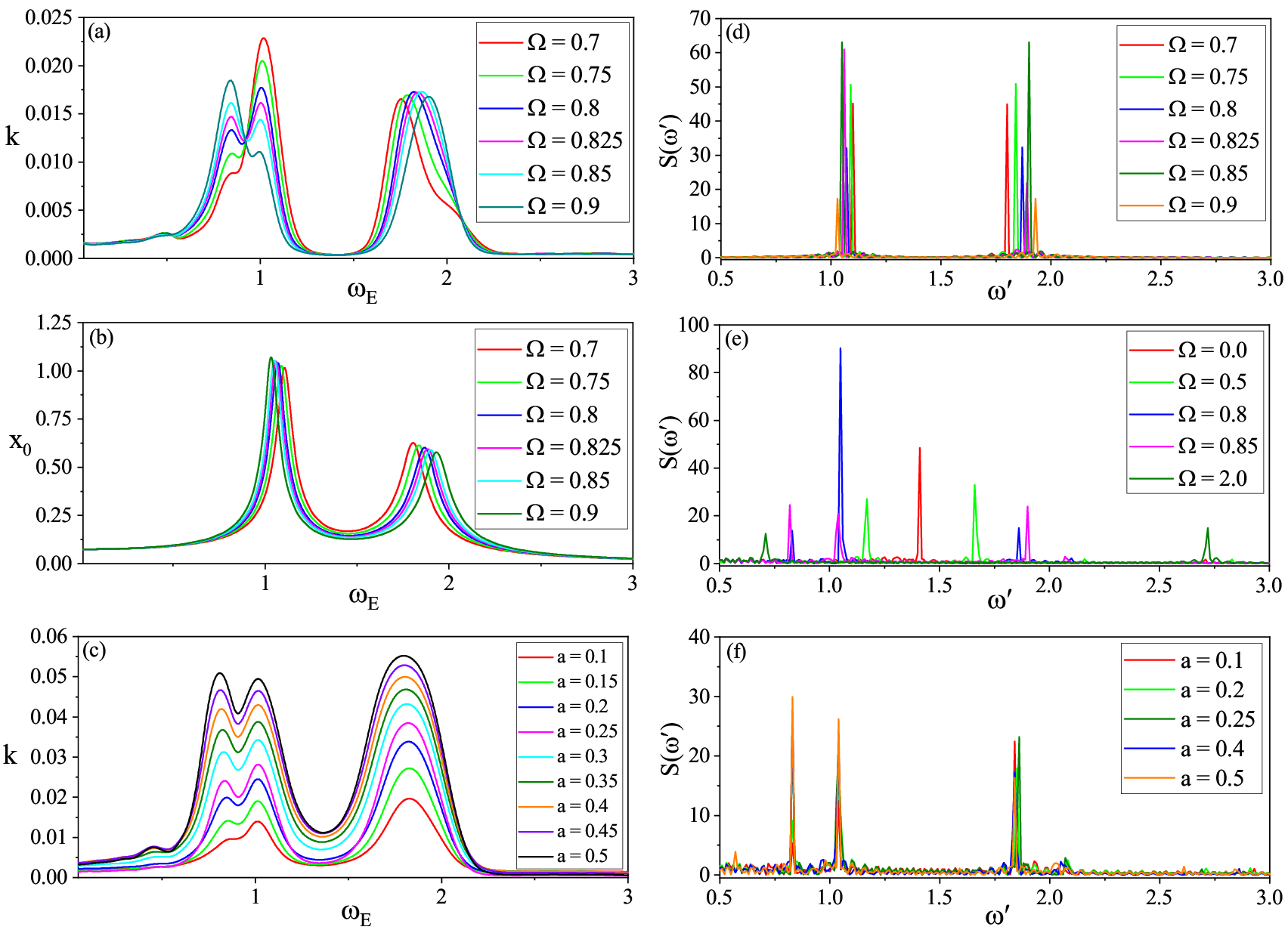}
\caption{(a) Plot of $k$ vs. $\omega_E$ for various values of $\Omega$ with the parameter set, $a = 0.25$, $b = 0.5$, $\omega_y^2 = 2.0$, $\gamma_a=0.1$, $E_{0x}=E_{0y}=0.15$ and $k_BT=0.04$ (b) $x_0$ vs. $\omega_E$ for various values of $\Omega$ with the parameter set,  $\omega_0^2 = .0$, $\omega_y^2 = 2.0$, $\gamma_a=0.1$, $E_{0x}=E_{0y}=0.15$ and $\gamma_a=0.1$. (c) Plot of  $k$ vs. $\omega_E$ for various values of $a$ with the parameter set,  $b = 0.5$, $\omega_y^2 = 2.0$, $\Omega_y^2 = 0.85$, $E_{0x}=E_{0y}=0.15$, $\gamma_a=0.1$ and $k_BT=0.04$. (d) Plot of 
$S(\omega^\prime)$ vs. $\omega^\prime$ for harmonic oscillator with $\omega_0^2 =2.0$ and $\omega_y^2 = 2.0$ (e) Plot of $S(\omega^\prime)$ vs. $\omega^\prime$ for the dynamics at the well of the bi-stable potential energy field with the parameter set, $a = 0.25$, $b = 0.5$ and $\omega_y^2 = 2.0$ (f) Same plot as in panel (e) for the parameter set, $b = 0.5$ and $\Omega_y^2 = 0.85$ and $\omega_y^2 = 2.0$ (Units are arbitrary).}
\label{fig.4}
\end{figure}

Careful inspection at intermediate strength of the magnetic field suggests an additional splitting of the spectrum around the lower resonating frequency ($\Omega_L$), Fig. \ref{fig.4}(a).
The amplitude, $x_0$, corresponding to the relevant parameter set is demonstrated in panel (b) of the same figure.
It implies that the additional splitting may be due to the anharmonic contribution from the potential energy field. To check this, we calculate the spectrum for different values of $a$ in Fig. \ref{fig.4}(c). Thus panel (c) clearly implies that the additional splitting is a signature of the non-linearity of the potential energy field. 
Quantitative analysis of these observations is very difficult since the relevant deterministic equations of motion are not solvable analytically. Then we solve the equations of motion (\ref{eq1}-\ref{eq2},\ref{eq4}-\ref{eq5}) numerically in the absence of thermal bath and driving field, and determine the Fourier transform of time-dependent position. The transformation is defined as

\begin{equation}\label{ft}
S(\omega^\prime)=2\int_0^\infty x(t) \cos \omega^\prime t dt   \;\;\;.
\end{equation}

\noindent
The result of the above integration for the equations of motion (\ref{eq10}-\ref{eq11}) with $\gamma=E_{0x}=E_{0y}=0$ is plotted in panel (d) of Fig. \ref{fig.4}. It corroborates panel (b) and implies that the additional splitting may be due to the anharmonicity of the potential energy field. The Fourier transformation corresponding to the dynamics at the well of the bi-stable potential is demonstrated in panels (e) and (f) of Fig. \ref{fig.4}. Panel (e) corroborates both the appearance and disappearance of the additional splitting. 
It signifies that if the frequency of the low frequency mode becomes relatively small, then the an-harmonic contribution from the potential energy field interferes significantly with it and renders an additional peak as well as splitting of the spectrum around $\Omega_L$. On further decreasing of the frequency of the mode
such that, it becomes close to the frequency of the non-linearity induced mode. Then the splitting may disappear as implied by panels (a) and (e)  in Fig. \ref{fig.4}. 
Finally, panel (e) corroborates the results of panel (c). Thus the additional splitting is a signature of non-linearity of the potential energy field.

\subsection{The asymmetry with respect to the height of peaks}

We are now in a position to consider the asymmetry with respect to peak heights. The relevant amplitudes at the resonance condition we obtain from the relation (\ref{eqr11}-\ref{eqr12}) as \cite{rahul}

\begin{equation}\label{eqr33}
x_0 \simeq \frac{q}{m}\frac{\sqrt{\left(\omega^2 - \omega_E^2\right)^2 \left(\gamma E_{0x} + 2 \Omega E_{0y} \right)^2+\omega_E^2 \left[\gamma\left(\gamma E_{0x} - \Omega E_{0y}\right) + 2 \Omega \left(\gamma E_{0y} + \Omega E_{0x}\right)\right]^2}}{\gamma^3 \omega_E^2+ 4\gamma \Omega^2 \omega_E^2}   \;\;\;,
\end{equation}

\noindent
and

\begin{equation}\label{eqr34}
y_0 \simeq \frac{q}{m}\frac{\sqrt{\left(\omega^2 - \omega_E^2\right)^2 \left(\gamma E_{0y} - 2 \Omega E_{0x} \right)^2+\omega_E^2 \left[\gamma \left(\gamma E_{0y} + \Omega E_{0x}\right) - 2 \Omega \left(\gamma E_{0x} - \Omega E_{0y}\right)\right]^2}}{\gamma^3 \omega_E^2 + 4\gamma \Omega^2 \omega_E^2}   \;\;\;.
\end{equation}

\noindent
In the absence of a magnetic field, the above equations reduce to

\begin{equation}\label{eqr33a}
x_0 = \frac{qE_{0x}}{m\gamma \omega_E}  \;\;\;,
\end{equation}

\noindent
and

\begin{equation}\label{eqr34a}
y_0 = \frac{qE_{0y}}{m\gamma \omega_E}  \;\;\;.
\end{equation}

\noindent
Thus, at the resonance condition, the finite amplitude for the harmonic oscillator in the presence of the energy dissipation decreases with an increase in the resonating frequency. The energy dissipation is 
strongly modulated by the magnetic field as implied by the denominator in Eqs. (\ref{eqr33}-\ref{eqr34}). \textcolor{blue}{Thus, for a given strength of the magnetic field, the amplitude as well as the rate constant at a higher resonating frequency may be lower than the other one as shown in Fig.~\ref{fig.1}. This figure implies that even at large $\Omega$, peak height corresponding to higher frequency may be vanishingly small. It is consisten with  Eqs.(\ref{eqr33}-\ref{eqr34}). At large $\Omega$, $\omega_R =\omega_E\simeq \Omega$ and then from the relations, (\ref{eqr33}-\ref{eqr34}) we have $x_0 \simeq y_0\simeq 1/\omega_E\simeq 1/\Omega$. In other words, at this regime, the value of resonating frequency corresponding to the right peak increases linearly with the cyclotron frequency.}

It is to be noted that there may be another reason for the asymmetric splitting of the spectrum concerning peak height. To address the other reason, we notice that in the presence of a magnetic field, the numerator of the amplitude function may depend on the damping strength. It is sharply contrasting to the harmonic oscillator case. Then the numerators in Eqs~.(\ref{eqr33}-\ref{eqr34}) may carry the signature of interference (as they depend on the frequency of the harmonic oscillator's, driving field, the damping strength, and the intensity of the applied field) between the driving components through velocity-dependent coupling.
Thus ,for a particular phase relationship between the driving components, the amplitude at a higher frequency may be greater than the other in the asymmetry splitting process. We will check this in the following subsection.

\textcolor{blue}{We now address two more important features of Fig.1(a). At zero magnetic field, the barrier crossing rate constant, $k=0.04$ at peak, and at different magnetic fields, its value decreases up to $k=0.003$ and less to $ k=0.001$. Peak height at low resonating frequency seems to be quite inconsistent according to panels (b) and (c). Then, in addition to the dynamical resonance, we would consider how the relevant Brownian motion, as well as the diffusion of the charged particle, is affected in the presence of the magnetic field. It was reported in 
Ref.\cite{filliger,abdoli1,abdoli}
that the Lorentz force slows down the escape dynamics via a trivial rescaling of the diffusion coefficient without affecting the exponential dependence on the barrier height. This may be the key reason for the decrease of peak height in the presence of MF.
Finally, one may notice from Fig.1(a) that there are two possibilities to cross the barrier with enhanced value in
comparison with another, but this value is 2-3 times less than at zero magnetic field. It may be due to the field-induced modulation of the resonating frequency of the dynamical system and the diffusion coefficient, respectively. Around the dynamical resonance zone, the value of the rate constant is enhanced that at zero magnetic field, but this value is 2-3 times less, as a signature of slowdown the escape dynamics via a trivial rescaling of the diffusion coefficient. In other words, suppression of the rate constant in the presence of a magnetic field around the driving frequency when the dynamical system experiences the energy activation through the resonance for zero magnetic field.}  

\textcolor{blue}{Before leaving this subsection, we note the following point. To avoid any confusion, it is to be noted here that the asymmetric spectrum for the rate constant (as demonstrated above with explanation) is not due to the asymmetric nature of the potential energy field as given by Eq.(\ref{eq5}). Therefore, one may expect that it would remain the same in the presence of the same cross fields
for the hopping of the particle along $y$-direction corresponding to the potential energy field,}
\textcolor{blue}{
\begin{equation}\label{eq5a}
V(x,y,z) = a y^4 - b y^2 +  (\omega_x^2 x^2 +\omega_z^2 z^2)/2   \; \; \;,
\end{equation}}
\noindent
\textcolor{blue}{with $\omega_x=\omega_y$. We check that Fig. \ref{fig.1} is reproduced for the equivalent parameter set. It is corroborated by the same relevant dynamics (as given Eqs. (\ref{eq10a}-\ref{eq11a}) around the bottom of the well. We also check that even the spectrum is qualitatively the same for the 
following symmetric potential energy field in the $x-y$ plane}
\textcolor{blue}{
\begin{equation}\label{eq5b}
V(x,y,z) = a(x^4+y^4) - b (x^2+y^2)+(\omega_x^2 x^2 +\omega_z^2 z^2)/2   \; \; \;.
\end{equation}}

\noindent
\textcolor{blue}{Although the relevant dynamics around the bottom of the well is the same (\ref{eq10a}-\ref{eq11a}) for this potential function but the contribution from the nonlinearity becomes double compared to the previous cases. The position of a peak and its height in Fig.1(a) may be modulated for the present case since the resonating frequency and the interference between driving components seem to be perturbed
by the enhanced contribution due to the anharmonicity.  But the nature of the spectrum is qualitatively the same as demonstrated in Fig.1(a). Thus it is independent of direction of hopping in the $x-y$ plane. Thus, the potential function with an unstable fixed point as given by Eq.(\ref{eq5}) is chosen to explore the essential feature of the hopping kinetics of a charged particle in the presence of an electromagnetic field.}

\subsection{Effect of the phase difference between the driving components on the asymmetric splitting process}

\begin{figure}[!htb]
\includegraphics[width=17cm,angle=0,clip]{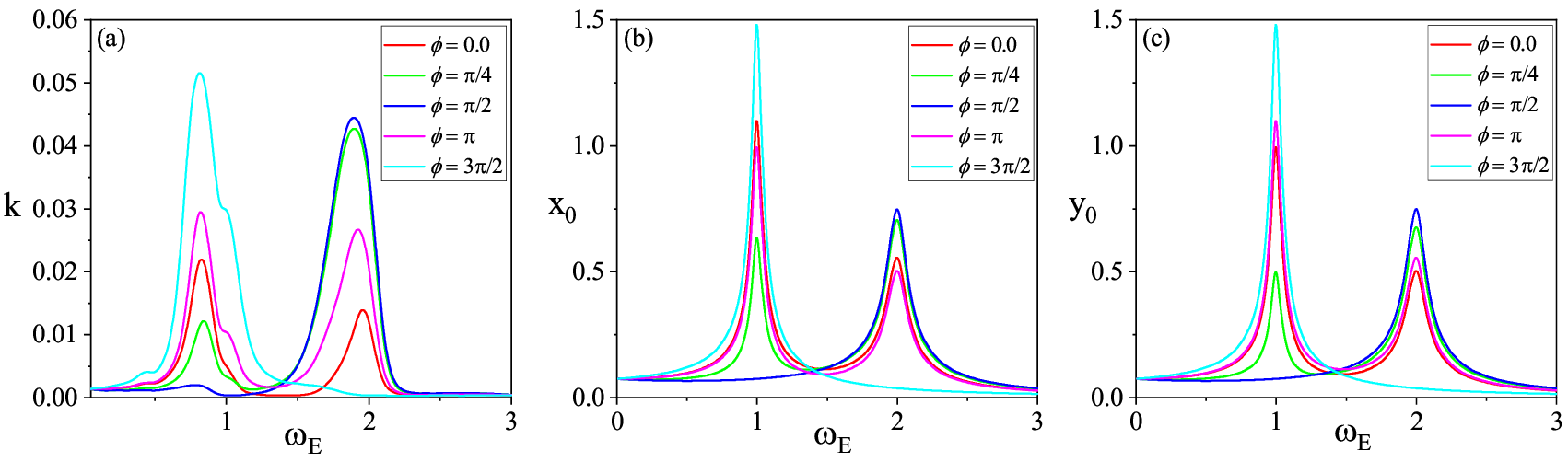}
\caption{Plot of (a) $k$ vs. $\omega_E$ at different values of $\phi$ for $a = 0.25$, $b = 0.5$, $\omega_y^2 = 2.0$ and $k_BT=0.04$; (b) $x_0$ vs. $\omega_E$ for different values of $\phi$ along with the relevant parameter set, $\omega_0^2 = \omega_y^2 = 2.0$. Common parameters are $\Omega = 1.0$, $\gamma = 0.1$ and $E_{0x} = E_{0y} = 0.15$. (Units are arbitrary).}
\label{fig.2}
\end{figure}

In Fig.~\ref{fig.2}(a), we demonstrate the effect of the phase difference between the driving components on the barrier crossing rate constant. It shows that for a given phase difference, the peak height at a higher resonating frequency may be greater than the other in the asymmetric splitting process. Even one of the peaks may disappear. Thus, the energy dissipation may be insignificant for a given phase difference to determine the peak height. This kind of unique asymmetric splitting process can be interpreted as a signature of interference between the driving components through the velocity-dependent coupling. \textcolor{blue}{Because the position of the peaks does not depend on the phase difference. It is expected since the relevant frequencies of the undriven dynamical system and the transient motion are the same for all the phase differences. Furthermore,} we check whether the dynamics around the bottom of the well can explain the asymmetric splitting process, we demonstrate the variation of the relevant amplitude with the driving frequency, Fig.~\ref{fig.2}(b). There is an excellent correspondence between the rate constant and the amplitude of the output signal. Finally, we will mention here that the order of peak heights in the spectrum of the rate constant for different values of the cyclotron frequency does not follow the same
with the amplitude (as shown in Fig.1) corresponding to the dynamics around the bottom of the relevant well. It may be due to the modulation of the interference through the anharmonicity of the potential energy field. 

\subsection{Dependence of the rate constant on the cyclotron frequency}

\begin{figure}[!htb]
\includegraphics[width=17cm,angle=0,clip]{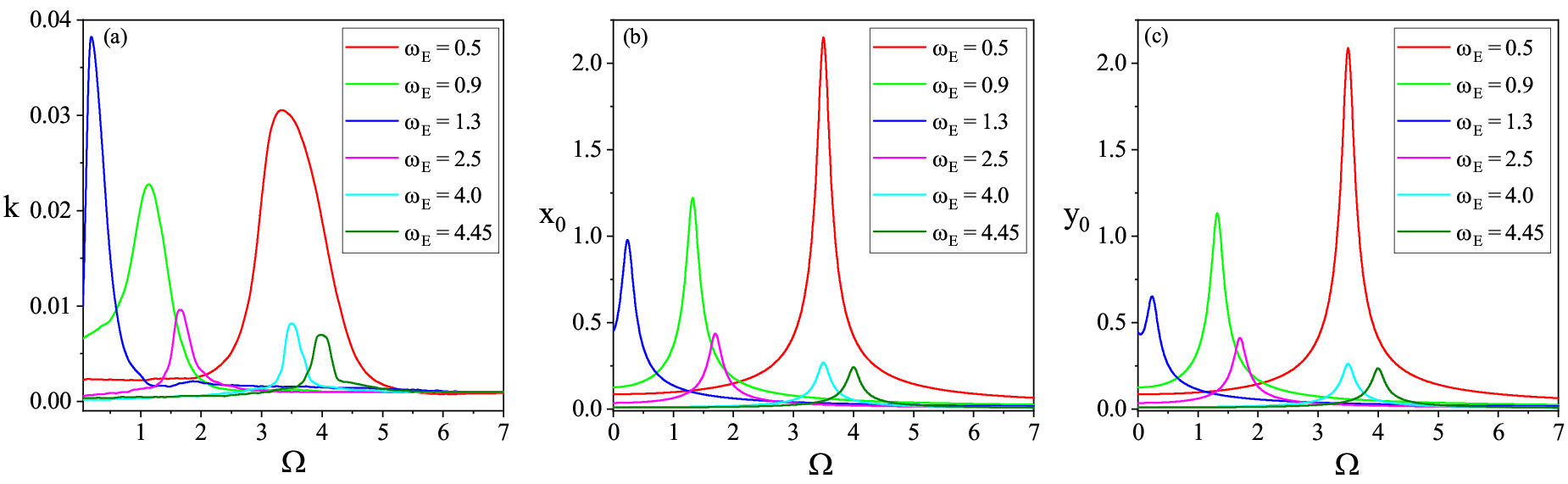}
\caption{Plot of (a) $k$ vs. $\Omega$ at different values of $\omega_E$ for $a = 0.25$, $b = 0.5$, $\omega_y^2 = 2.0$ and $k_BT=0.04$; (b) $x_0$ and  $\Omega$ for different values of $\omega_E$ along with the relevant parameter set, $\omega_0^2 = \omega_y^2 = 2.0$. Common parameters are $\gamma = 0.1$ and $E_{0x} = E_{0y} = 0.15$. (Units are arbitrary).}
\label{fig.5}
\end{figure}

In Fig. \ref{fig.5}, we demonstrate how the rate constant depends on the strength of the applied magnetic field for a given periodic electric field. It shows that there is a critical magnetic field strength at which the rate constant is maximum. One can explain this observation in the following way. The driving field is searching for a relevant frequency from the dynamical system (which experiences an applied magnetic field) to be effective in the energy transfer process. For a certain strength of the applied magnetic field, one of the frequencies of the dynamical system may match the driving frequency, which results in the appearance of a maximum. One can determine the position of the maximum 
approximately, considering the steady state dynamics around the bottom of the well \cite{rahul}. The $\omega_L$ and 
$\omega_R$ in Eqs.(\ref{eqr31}-\ref{eqr32}) are the relevant roots of the following equation,

\begin{equation}\label{eqr30}
\left(\omega^2 - \omega_E^2\right)^2 - \Omega^2 \omega_E^2 = 0   \;\;\;.
\end{equation}

\noindent
Then one may find that the position of the maximum at

\begin{equation}\label{eqr30a}
\Omega_P \simeq \frac{\big| \omega^2 - \omega_E^2 \big|}{\omega_E}   \;\;\;.
\end{equation}

\noindent
\textcolor{blue}{It is to be noted here that at large $\Omega$, the resonating driving frequency for the right peak, $\omega_E=\omega_R>>\omega$. Under these conditions the above equation becomes}
\textcolor{blue}{
\begin{equation}\label{eqr30b}
\Omega_P \simeq \omega_E=\omega_R   \;\;\;.
\end{equation}}

\noindent
\textcolor{blue}{Thus if a peak appears at small $\Omega$ then $\Omega_P\neq\omega_E$ and it is consistent with Fig.\ref{fig.5}. In other words, for comparatively large driving frequency ($\omega_E$), the peak may appear at $\Omega=\Omega_P \simeq \omega_E=\omega_R$.  This is also consistent with Fig.\ref{fig.5} with  $\omega_E=    $ and $ $, respectively. Thus at this regime, the value of cyclotron frequency corresponding to the position of a peak increases linearly with the driving frequency. It is corroborated by Eqs. (\ref{eqr33}-\ref{eqr34}) as discussed after Eq.(\ref{eqr34a}). Finally,} the value of $\Omega_R$ based on the Eqs. (\ref{eqr30a}-\ref{eqr30b})) is compared with the exact result in Table \ref{tab.2}. It shows a fair agreement between the theory and the numerical experiment.

\begin{table}[ht]
\caption{Comparison between theoretically calculated peak position and the exact result}
\begin{center}
\begin{tabular}{|c|c|c|}
\hline
Value of &
\multicolumn{2}{|c|}{Peak at $\Omega$} \\
\cline{2-3}
$\omega_E$ & Theoretical & Exact \\ 
\hline
0.5 & 3.500 & 3.350 \\
\hline
0.9 & 1.320 & 1.149 \\
\hline
1.3 & 0.238 & 0.200 \\
\hline
2.5 & 1.700 & 1.649 \\
\hline
4.0 & 3.500 & 3.500 \\
\hline
4.45 & 4.000 & 4.000 \\
\hline
\end{tabular}
\end{center}
\label{tab.2}
\end{table}

\section{Barrier crossing dynamics in the presence of electric and magnetic fields at arbitrary directions}


In this section, we consider a magnetic field (${\bf B} = (B_x, B_y, B_z)$) atan  arbitrary direction. Then the relevant equations of motion can be read as

\begin{equation}\label{anis2}
 \ddot{x} = -4ax^3+2bx^2 +  \Omega_z \dot{y} - \Omega_y \dot{z}+ q E_{0x} \cos(\omega_E t - \phi_x)-  \gamma \dot{x}+f_x(t)   \;\;\;,
\end{equation}

\begin{equation}\label{anis3}
\ddot{y} = - \omega_y^2 y  +  \Omega_x \dot{z} -  \Omega_z \dot{x}+ q E_{0y} \cos(\omega_E t - \phi_y)-  \gamma \dot{y}+f_y(t)   \;\;\;,
\end{equation}

\begin{equation}\label{anis4}
\ddot{z} = -  \omega_z^2 z  + \Omega_y \dot{x} -  \Omega_x \dot{y}+ q E_{0z} \cos(\omega_E t - \phi_z)-  \gamma \dot{z}+f_z(t)   \;\;\;.
\end{equation}

\noindent
Here we have used $\Omega_x = q B_x/m, \Omega_y = q B_y/m$ and $\Omega_x = q B_z/m$.  The term after the magnetic force on the right-hand side
of each of the above equations is due to the following time-dependent periodic electric field,

\begin{equation}\label{anis1}
{\bf E} = \hat{i} E_{0x} \cos(\omega_E t - \phi_x) + \hat{j} E_{0y} \cos(\omega_E t - \phi_y) + \hat{k} E_{0z} \cos(\omega_E t - \phi_z)   \;\;\;
\end{equation}

\begin{figure}[!htb]
\includegraphics[width=16.5cm,angle=0,clip]{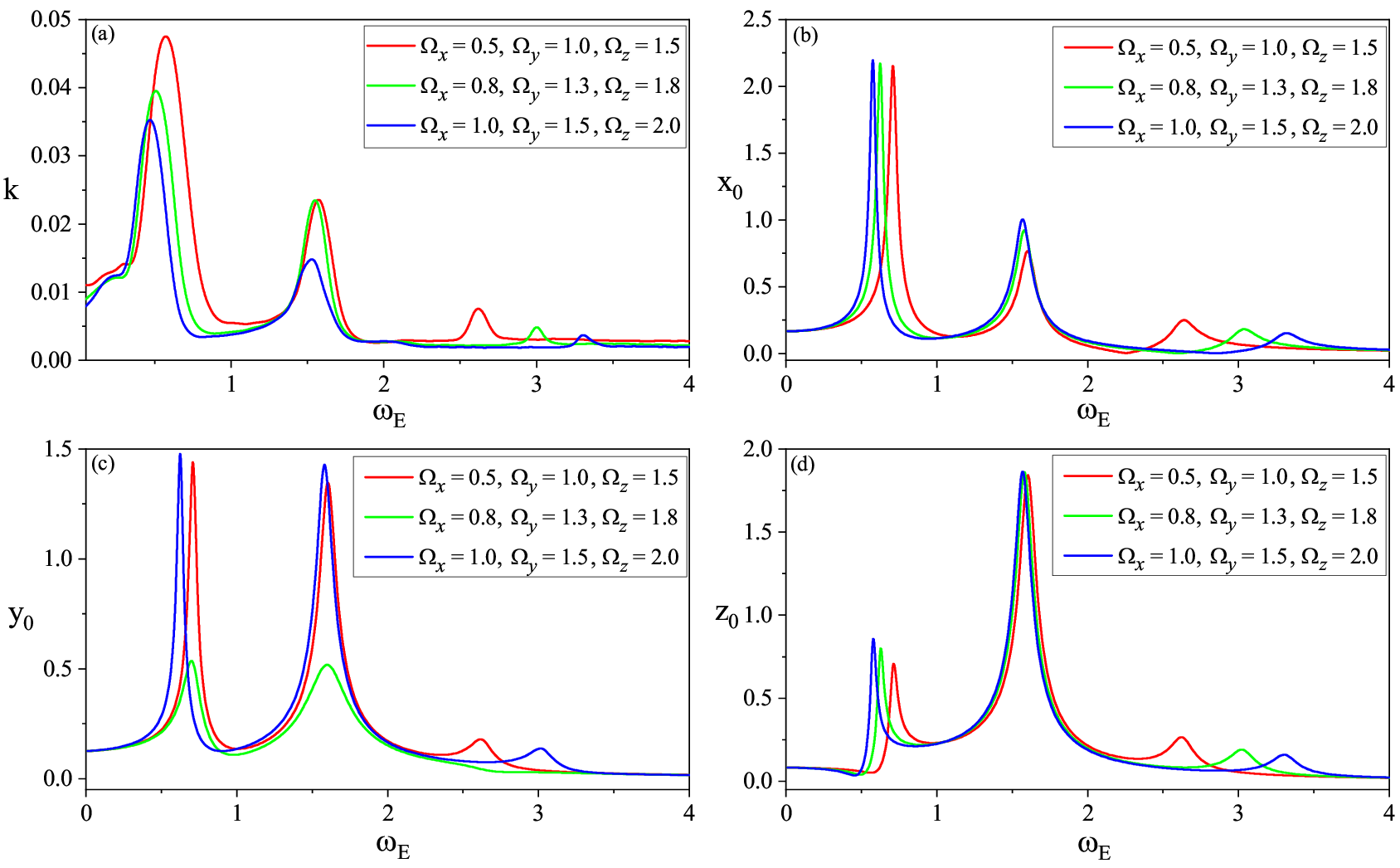}
\caption{Plot of (a) $k$ vs. $\omega_E$ at various values of $\Omega_x$, $\Omega_y$ and $\Omega_z$ with parameters, $a = 0.14$, $b = 0.375$, $\omega_y^2 = 2.0$, $\omega_z^2 = 3.0$ and $k_B T=0.04$; (b), (c), (d) $x_0$, $y_0$ and $z_0$ vs. $\omega_E$ for various values of $\Omega_x$, $\Omega_y$ and $\Omega_z$ along with the parameter set, $\omega_x^2 = 1.5$, $\omega_y^2 = 2.0$, $\omega_z^2 = 3.0$. Common parameters are $\gamma = 0.1$ and $E_{0x} = E_{0y} = 0.25$. (Units are arbitrary).}
\label{fig.6}
\end{figure}

We now demonstrate spectrum for the rate constant corresponding to the equations (\ref{anis2}-\ref{anis4}) of motion in Fig~.\ref{fig.6}. It shows that an additional peak appears between the two peaks compared to Fig.1(a). The additional peak is not so sensitive to the direction of the applied MF. \textcolor{blue}{It is to be noted here that it is difficult to explain these observations considering transient motion around the bottom of the left well as given in the previous section. Then we calculate $S(\omega^\prime)$ numerically for the following equations of motion,}

\textcolor{blue}{
\begin{equation}\label{anis2}
\ddot{x} = -4a x^3 + 2bx + \Omega_z \dot{y} - \Omega_y \dot{z}- \gamma \dot{x}   \;\;\;,
\end{equation}}

\textcolor{blue}{
\begin{equation}\label{anis3}
\ddot{y} = - \omega_y^2 y  + \Omega_x \dot{z} -  \Omega_z \dot{x}-\gamma \dot{y}   \;\;\;,
\end{equation}}

\noindent
and
\textcolor{blue}{
\begin{equation}\label{anis4}
\ddot{z} = -  \omega_z^2 z  + \Omega_y \dot{x} -  \Omega_x \dot{y}-\gamma \dot{z}  \;\;\;.
\end{equation}}
\textcolor{blue}{
\begin{figure}[!htb]
\includegraphics[width=16.5cm,angle=0,clip]{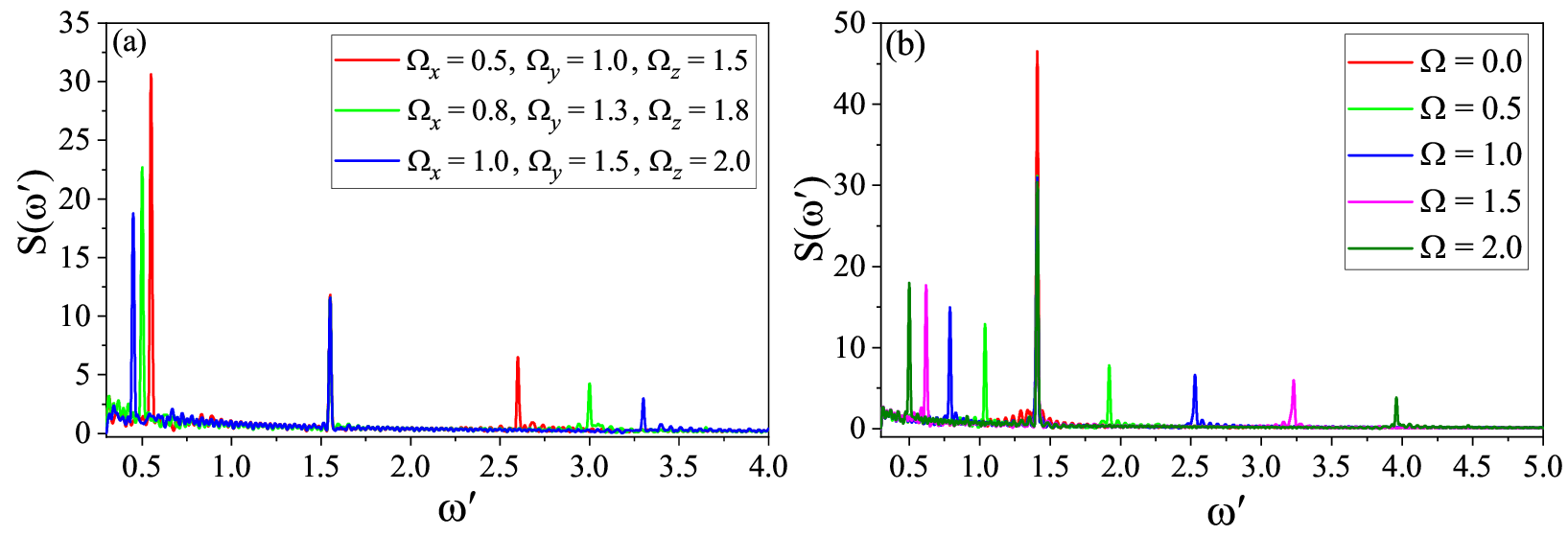}
\caption{(a) Plot of $S(\omega^\prime)$ vs. $\omega^\prime$ for the dynamics at the well of the bi-stable potential energy field with the parameter set, $a = 0.14$, $b = 0.375$, $\omega_y^2 = 2.0$ and $\omega_z^2 = 3.0$; (b) Same plot as in panel (a) for the parameter set, $a = 0.25$, $b = 0.5$ and $\omega_y^2 = \omega_z^2 = 2.0$. (Units are arbitrary).}
\label{fig.6a}
\end{figure}}

\noindent
\textcolor{blue}{$S(\omega^\prime)$ is demonstrated in panel (a) of Fig~.\ref{fig.6a}. It implies that the oscillating motion is composed of three frequencies and therefore we observe three peaks in the spectrum with the rate constant corresponding to the energy activation through the dynamical resonance
for the electric field driven dynamical system. Thus it is apparent here that in addition to the asymmetric splitting of the frequency as like to the two-dimensional motion, a state of motion may appear at which the rotational motion due to the potential energy field is neither favored nor opposed by the magnetic field induced rotational motion. In other words, a state of motion may be possible when the net magnetic force on the particle is vanishingly small as a result of interference between the components of the same as implied in the equations of motion. This is corroborated by the two-dimensional motion where this kind of interference is not possible.} 

However, considering the steady state dynamics around the bottom of the well, one may account for the peaks. We calculate the amplitude of the driven damped three-dimensional anisotropic harmonic oscillator as like as in the previous section. Following Ref. \cite{rahul}, the components of the amplitude are given by

\begin{equation}\label{anis8}
x_0 =  \frac{\sqrt{A_1^2 + A_2^2}}{D_1}  \;\;\;,
\end{equation}

\begin{equation}\label{anis9}
y_0 = \frac{\sqrt{B_1^2 + B_2^2}}{D_1}   \;\;\;,
\end{equation}

\begin{equation}\label{anis10}
z_0 = \frac{\sqrt{C_1^2 + C_2^2}}{D_2}   \;\;\;.
\end{equation}

\noindent
$A_1, A_2, B_1, B_2, C_1, C_2, D_1$ and $D_2$ which appear in Esq.(\ref{anis8}-\ref{anis10}) are defined in Appendix A. Calculating $x_0$, $y_0$ and $z_0$ for the same parameter set for Fig.~\ref{fig.6}(a), we demonstrate their dependence on the driving frequency in panels (b)-(d) in the same figure, which corroborate strongly with panel (a). However, it is very difficult to determine the location of peaks for this case. One may find the reason in Ref. \cite{rahul}. Then we consider the dynamics for the simplest case, $\omega_x = \omega_y = \omega_z = \omega$, $\Omega_x = \Omega_y = \Omega_z = \Omega$ and $\phi_x=\phi_y=\phi_z=0$. For this case  
the locations of left, middle, and right  peaks \cite{rahul}
are at

\begin{equation}\label{anis60}
\omega_L = \sqrt{\omega^2 + \frac{\Omega^2}{2}} - \sqrt{\frac{\Omega^2}{2}}   \;\;\;,
\end{equation}

\begin{equation}\label{anis59}
\omega_m = \omega   \;\;\;
\end{equation}

\noindent
and

\begin{equation}\label{anis61}
\omega_R = \sqrt{\omega^2 + \frac{\Omega^2}{2}} + \sqrt{\frac{\Omega^2}{2}}   \;\;\;.
\end{equation}

\begin{figure}[!htb]
\includegraphics[width=16.5cm,angle=0,clip]{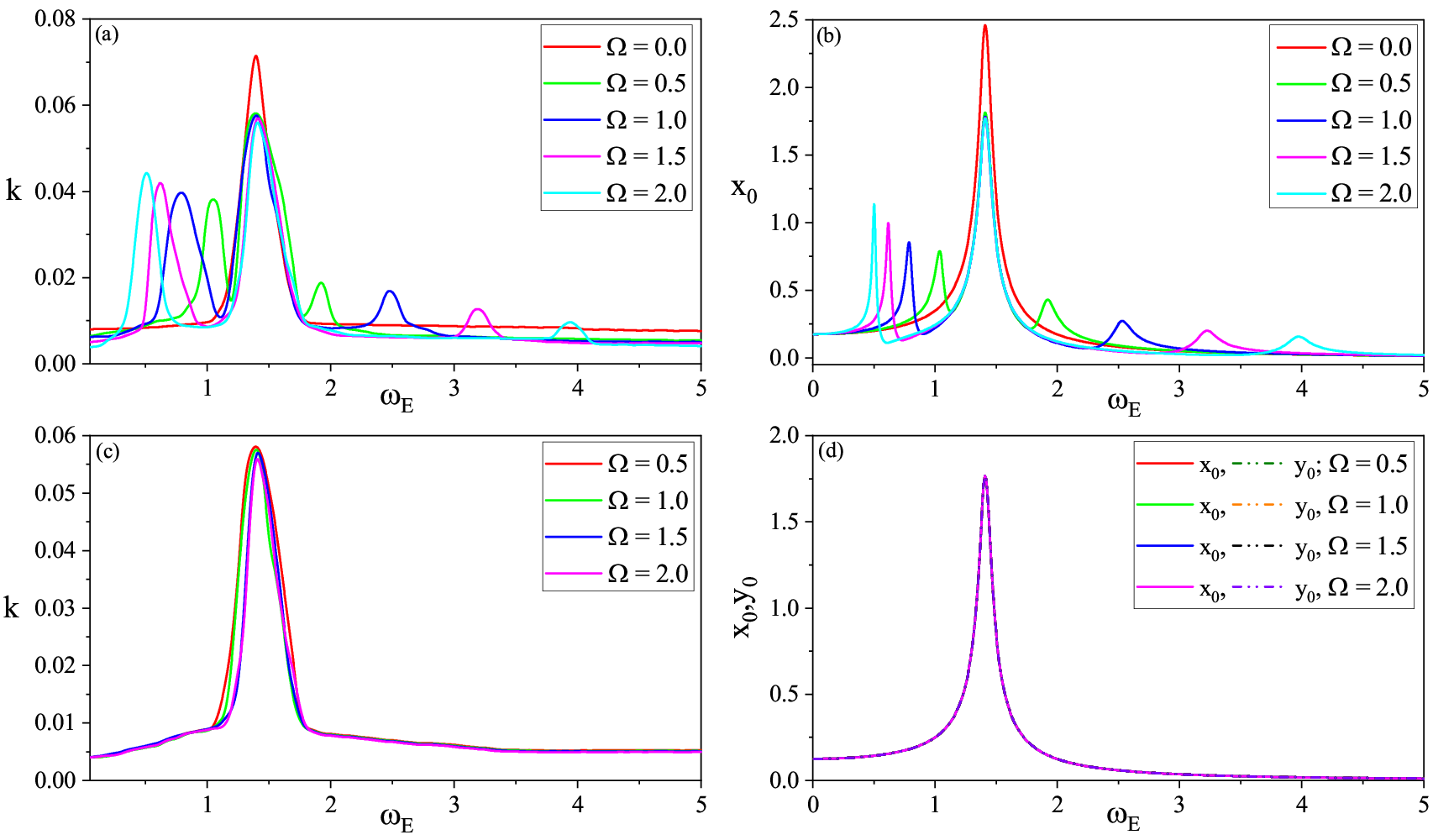}
\caption{Plot of (a) $k$ vs. $\omega_E$ and (b) $x_0$ vs. $\omega_E$ for various values of $\Omega$ with $E_{0x} = 0.55$, $E_{0y} = 0.45$, $E_{0z} = 0.35$. Plot of (c) $k$ vs. $\omega_E$ and (d) $x_0$, $y_0$ vs. $\omega_E$ for various values of $\Omega$ with $E_{0x} = E_{0y} = E_{0z} = 0.55$. Common parameters are $a = 0.14$, $b = 0.375$, $\omega_y^2 = 2.0$, $\omega_z^2 = 3.0$, $k_BT=0.04$ for (a) and (c); $\omega_x^2 = 1.5$, $\omega_y^2 = 2.0$, $\omega_z^2 = 3.0$ for (b) and (d); $\gamma = 0.1$ for all. (Units are arbitrary).}
\label{fig.7}
\end{figure}

To check these results, we calculate the rate constant for the relevant parameter set and demonstrate the same in panel (a) of Fig. \ref{fig.7}. It shows very good agreement between theory and numerical experiments. \textcolor{blue}{This result is corroborated by panel (b) of Fig.~\ref{fig.6}(a) and panel (b) of Fig.\ref{fig.7}, respectively. It is to be noted here that for $E_{0x} = E_{0y} = E_{0z} = E_0$  panel (a) in Fig.\ref{fig.7} becomes panel (c) in the same figure. It is to be noted here that the undriven dynamical system is the same for both panels (a) and (c) of  Fig. \ref{fig.7}. Then panel (b) of Fig.~\ref{fig.6}(a) implies that disappearance of the two peaks is certainly a signature of interference between the driving components through the velocity-dependent coupling among the equations of motion.
Further more, considering Eqs.(\ref{anis8}-\ref{anis10}) one may account for panel (c).} 
For the condition corresponding to it, the components of the amplitude become

\begin{equation}\label{eq19}
x_0 = \frac{q E_0 N}{m D^\prime H_0 \left[\left(\omega^2 - \omega_E^2\right)^2 - \gamma^2 \omega_E^2\right]^{1/2}}  \;\;\;,
\end{equation}

\begin{equation}\label{eq20}
y_0 = z_0 =\frac{q E_0 N^\prime}{m D^\prime H_0 \left[\left(\omega^2 - \omega_E^2\right)^2 - \gamma^2 \omega_E^2\right]^{1/2}}   \;\;\;.
\end{equation}

\noindent
where $N = \big(H_0 - H_2\big) \big(H_0^2 + H_1 H_2\big) + \big(H_0 + H_1\big) \big(H_2^2 + H_0 H_1\big)$,
$N^\prime = \big(H_0 + H_1\big) \big(H_0^2 + H_1 H_2\big) + \big(H_0 - H_2\big) \big(H_1^2 - H_0 H_2\big)$,
$D = \big(H_0^2 + H_1 H_2\big)^2 - \big(H_1^2 - H_0 H_2\big) \big(H_2^2 + H_0 H_1\big) = D^\prime H_0 \big(H_0 - H_1 + H_2\big)$,
$D^\prime = \big\{H_0 \big(H_0 - H_1 + H_2\big) + 3 \Omega^4 \omega_E^4\big\}$, 
$H_0 = \big(\omega^2 - \omega_E^2\big)^2 - \big(2 \Omega^2 - \gamma^2\big) \omega_E^2$, $H_1 = \Omega \omega_E^2 \big(2 \gamma - \Omega\big)$ and $H_2 = \Omega \omega_E^2 \big(2 \gamma + \Omega\big)$.
$x_0$ and $y_0$ are demonstrated in panel (d). Thus, the dynamical resonance implies that only one peak appears if the bottom of the well mimics an isotropic harmonic oscillator. Following Ref.\cite{rahul} one can determine the location of the peak from Eqs.(\ref{eq19}-\ref{eq20}) as

\begin{equation}\label{eq21}
\omega_m = \omega   \;\;\;.
\end{equation}

\noindent
Thus, the approximate calculation agrees well with  the numerical experiment. \textcolor{blue}{In other words, the steady state dynamics implies that the only peak is due to the interference among the driving components through the velocity-dependent coupling in the equations of motion.}

\begin{figure}[!htb]
\includegraphics[width=16.5cm,angle=0,clip]{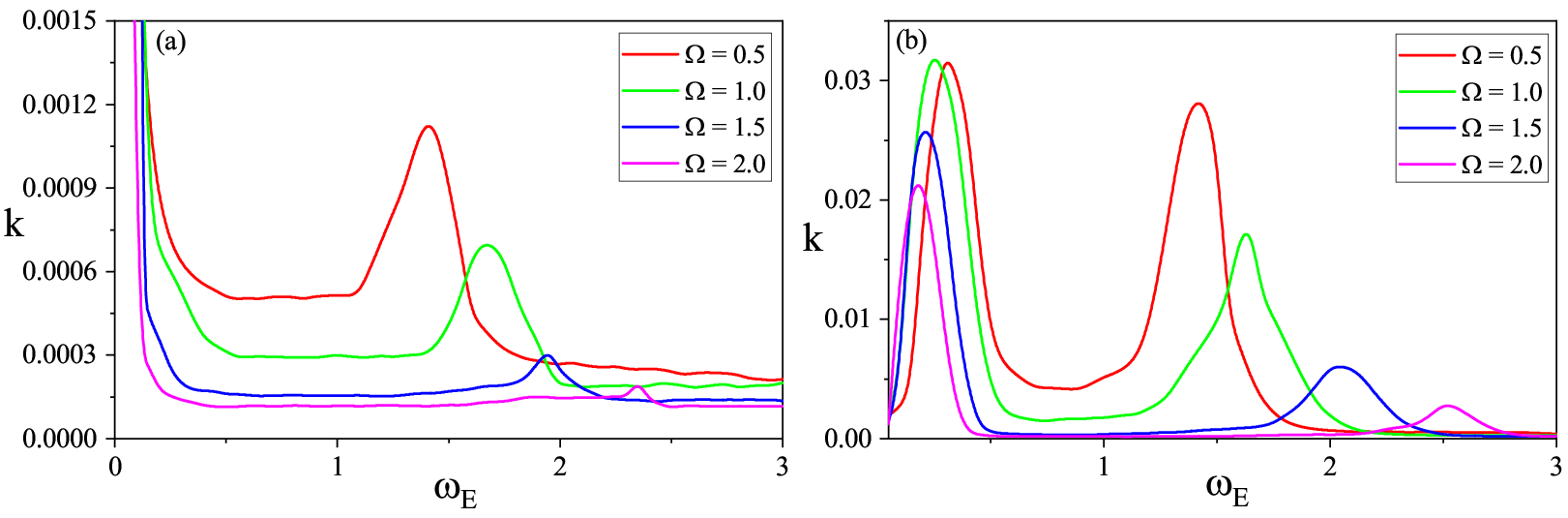}
\caption{Plot of $k$ vs. $\omega_E$ at different values of $\Omega$ for (a) $a = 0.25$, $b = 0.5$, $\omega_y^2 = 0.0$ ($\omega_0^2 = 2.0$, $\omega_y^2 = 0.0$), (b) $a = 0.25$, $b = 0.5$, $\omega_y^2 = 0.1$ ($\omega_0^2 = 2.0$, $\omega_y^2 = 0.1$; $\omega_0 \gg \omega_y$). Common parameters are $\gamma = 0.1$, $k_BT=0.04$ and $E_{0x} = E_{0y} = 0.15$. (Units are arbitrary).}
\label{fig.8}
\end{figure}

Finally, the present section may be helpful in considering a relevant case of the previous section, the frequencies at the bottom of the double well potential are not the same, i.e., $\omega_0\neq\omega_y$. The rate constant corresponding to this situation is calculated and demonstrated in Fig.~\ref{fig.8}. It shows that for this case, two peaks also appear except for the condition, $\omega_y=0$. \textcolor{blue}{Earlier discussion suggests that two peaks are expected for $\omega_y\neq 0$.
Following Ref.~\cite{rahul}, the location of the peaks can be determined approximately for $\omega_0<<\omega_y$ or $\omega_0>>\omega_y$. 
Since the resonance phenomenon may appear at $\gamma\rightarrow0$
Then the relevant amplitude of the output signal may be maximum around the
following condition}
\textcolor{blue}{
\begin{equation}
\left(\omega_{0}^{2}-\omega_{E}^{2}\right)\left(\omega_{y}^{2}-\omega_{E}^{2}\right)-\Omega^{2}\omega_{E}^{2}=0\;\;\;.\label{df12a}
\end{equation}}

\noindent \textcolor{blue}{If $\omega_{y}=0$, then the solution of the above can be
$\omega_{E}=\sqrt{\omega_{0}^{2}+\Omega^{2}}$ and
$\omega_{E}=0$. This condition implies that for
the one-dimensional harmonic oscillator in the presence of a magnetic
field (which is perpendicular to the direction of the oscillator) only one resonance peak may appear. In other words, for the electric field driven barrier crossing dynamics,
a peak may appear at the driving frequency, $\omega_{E}=\sqrt{\omega_{0}^{2}+\Omega^{2}}$ in the spectrum for the rate constant as shown in panel (c) of Fig.~\ref{fig.8}.  To avoid any confusion, we would
mention here that the peak at $\omega_{E}=0$ 
corresponds to the diverging motion in the presence of a constant
electric field. One may verify it easily, considering the relevant
equations of motion for the dynamics around the bottom of the left well.
However, from the above equation,n we have 
$\omega_0 \ll \omega_y$, the relevant peaks may appear around the driving frequencies, $\omega_R \simeq \sqrt{\omega_0^2 + \Omega^2}$ and $\omega_L \simeq \frac{\omega_0 \omega_y}{\sqrt{\omega_0^2 + \Omega^2}}$, respectively as shown in panel () of Fig.~\ref{fig.8}. Similarly, for the other condition, the peaks may appear around the driving frequencies, $\omega_R \simeq \sqrt{\omega_y^2 + \Omega^2}$ and $\omega_L \simeq \frac{\omega_0 \omega_y}{\sqrt{\omega_y^2 + \Omega^2}}$, respectively.  Panel () of Fig.~\ref{fig.8} agrees well with the theoretical prediction.}

Before leaving this section, we would like to mention that for the present case, the non-linearity induced additional splitting also occurs, and the effect of phase difference among the driving components on the barrier crossing rate constant is similar to that of Fig. \ref{fig.2}.

\section{Conclusion}

We have investigated the barrier crossing dynamics of a time-dependent periodic electric field-driven Brownian particle in the presence of a constant magnetic field. It includes the following major points.

\noindent
(i) For the cross fields with low or high values of the cyclotron frequency, asymmetric splitting of the spectrum for the rate constant occurs with two peaks. Anharmonicity-induced additional splitting may appear around the lower resonating frequency at intermediate strength of the applied magnetic field. 

\noindent
(ii) If the fields are not perpendicular to each other, then one additional peak may appear between the two peaks.
The position of the additional peak may not be sensitive to the applied magnetic field strength. For the dynamics around the bottom of the well (similar to the Lorentz force-driven isotropic harmonic oscillator), the position of the middle peak is independent of the strength of the field, whose all components are the same. 

\noindent
(iii) In some cases, only one peak appears even in the presence of a magnetic field. It may be a signature of interference among the components of the motion.

\noindent
(iv) For a given time-dependent periodic electric field, there is only one peak in the spectrum for the rate constant as a function of the strength of the time-independent magnetic field.

\noindent 
(v) We have analyzed most of these observations, considering the dynamics around the 
stable fixed point. Exploration of the functional form of the rate constant is in progress.

\noindent
(vi) The present study might be useful for understanding the tuning of the conductivity of a solid electrolyte in the presence of an electromagnetic field. The tuning of the conductivity by a physical method instead of changing the chemical composition of a solid electrolyte is challenging in recent technology. Other applications may be in areas such as electromagnetic field-induced modulation of (a) thermally activated tunneling ionization, (b) thermally stimulated ionization, etc.

\vspace{0.5cm}

\noindent
{\bf Author's contributions}

\noindent
Major contributions are from L. R. R. Biswas and B. C. Bag.

\noindent
{\bf Acknowledgment}

\noindent
Thanks are due to the DST-INSPIRE, Council of Scientific and Industrial Research (CSIR), and the University Grants Commission (UGC), Government of India, for partial financial support.
We thank Mousumi Biswas for reading our manuscript carefully.

\noindent
{\bf DATA AVAILABILITY}

\noindent
The data that support the findings of this study are available within the article.

\appendix
\section{Definition of relevant quantities which appear in Eqs.(\ref{anis8}-\ref{anis10})}

\begin{equation}\label{anis14}
A_1 = \left(H_c H_x - H_z H_2\right) \left(H_b H_c + H_3 H_6\right) + \left(H_c H_y + H_z H_3\right) \left(H_c H_1 + H_2 H_6\right)   \;\;\;,
\end{equation}

\begin{equation}\label{anis15}
A_2 = \left(H_c H_x^\prime - H_z^\prime H_2\right) \left(H_b H_c + H_3 H_6\right) + \left(H_c H_y^\prime + H_z^\prime H_3\right) \left(H_c H_1 + H_2 H_6\right)   \;\;\;,
\end{equation}

\begin{equation}\label{anis16}
B_1 = \left(H_c H_y + H_z H_3\right) \left(H_a H_c + H_2 H_5\right) - \left(H_c H_x - H_z H_2\right) \left(H_c H_4 - H_3 H_5\right)   \;\;\;,
\end{equation}

\begin{equation}\label{anis17}
B_2 = \left(H_c H_y^\prime + H_z^\prime H_3\right) \left(H_a H_c + H_2 H_5\right) - \left(H_c H_x^\prime - H_z^\prime H_2\right) \left(H_c H_4 - H_3 H_5\right)   \;\;\;,
\end{equation}

\begin{equation}\label{anis18}
C_1 = \left(H_b H_z - H_y H_6\right) \left(H_a H_b + H_1 H_4\right) + \left(H_b H_x + H_y H_1\right) \left(H_b H_5 + H_4 H_6\right)   \;\;\;,
\end{equation}

\begin{equation}\label{anis19}
C_2 = \left(H_b H_z^\prime - H_y^\prime H_6\right) \left(H_a H_b + H_1 H_4\right) + \left(H_b H_x^\prime + H_y H_1\right) \left(H_b H_5 + H_4 H_6\right)   \;\;\;,
\end{equation}

\begin{equation}\label{anis20}
D_1 = \left(H_a H_c + H_2 H_5\right) \left(H_b H_c + H_3 H_6\right) + \left(H_c H_1 + H_2 H_6\right) \left(H_c H_4 - H_3 H_5\right)   \;\;\;.
\end{equation}

\begin{equation}\label{anis21}
D_2 = \left(H_a H_b + H_1 H_4\right) \left(H_b H_c + H_3 H_6\right) + \left(H_b H_5 + H_4 H_6\right) \left(H_b H_2 - H_1 H_3\right)   \;\;\;,
\end{equation}

\noindent
with

\begin{eqnarray}\label{anis22}
H_a &=& \left(\omega_x^2 - \omega_E^2\right)^2 \left(\omega_y^2 - \omega_E^2\right) \left(\omega_z^2 - \omega_E^2\right) + \Big\{\gamma^2 \left(\omega_y^2 - \omega_E^2\right) \left(\omega_z^2 - \omega_E^2\right)   \nonumber \\ \nonumber \\
&-& \Omega_y^2 \left(\omega_x^2 - \omega_E^2\right) \left(\omega_y^2 - \omega_E^2\right) - \Omega_z^2 \left(\omega_z^2 - \omega_E^2\right) \left(\omega_x^2 - \omega_E^2\right)\Big\} \omega_E^2   \;\;\;,
\end{eqnarray}

\begin{eqnarray}\label{anis23}
H_b &=& \left(\omega_x^2 - \omega_E^2\right) \left(\omega_y^2 - \omega_E^2\right)^2 \left(\omega_z^2 - \omega_E^2\right) + \Big\{\gamma^2 \left(\omega_z^2 - \omega_E^2\right) \left(\omega_x^2 - \omega_E^2\right)   \nonumber \\ \nonumber \\
&-& \Omega_z^2 \left(\omega_y^2 - \omega_E^2\right) \left(\omega_z^2 - \omega_E^2\right) - \Omega_x^2 \left(\omega_x^2 - \omega_E^2\right) \left(\omega_y^2 - \omega_E^2\right)\Big\} \omega_E^2   \;\;\;,
\end{eqnarray}

\begin{eqnarray}\label{anis24}
H_c &=& \left(\omega_x^2 - \omega_E^2\right) \left(\omega_y^2 - \omega_E^2\right) \left(\omega_z^2 - \omega_E^2\right)^2 + \Big\{\gamma^2 \left(\omega_x^2 - \omega_E^2\right) \left(\omega_y^2 - \omega_E^2\right)   \nonumber \\ \nonumber \\
&-& \Omega_x^2 \left(\omega_z^2 - \omega_E^2\right) \left(\omega_x^2 - \omega_E^2\right) - \Omega_y^2 \left(\omega_y^2 - \omega_E^2\right) \left(\omega_z^2 - \omega_E^2\right)\Big\} \omega_E^2   \;\;\;,
\end{eqnarray}

\begin{eqnarray}\label{anis25}
H_x &=& \frac{q}{m} \Big\{E_{0x} \left(\omega_x^2 - \omega_E^2\right) \left(\omega_y^2 - \omega_E^2\right) \left(\omega_z^2 - \omega_E^2\right) \cos \phi_x   \nonumber \\ \nonumber \\
&-& E_{0x} \gamma \omega_E \left(\omega_y^2 - \omega_E^2\right) \left(\omega_z^2 - \omega_E^2\right) \sin \phi_x   \nonumber \\ \nonumber \\
&+& E_{0y} \Omega_z \omega_E \left(\omega_z^2 - \omega_E^2\right) \left(\omega_x^2 - \omega_E^2\right) \sin \phi_y   \nonumber \\ \nonumber \\
&-& E_{0z} \Omega_y \omega_E \left(\omega_x^2 - \omega_E^2\right) \left(\omega_y^2 - \omega_E^2\right) \sin \phi_z\Big\}   \;\;\;,
\end{eqnarray}

\begin{eqnarray}\label{anis25}
H_y &=& \frac{q}{m} \Big\{E_{0y} \left(\omega_x^2 - \omega_E^2\right) \left(\omega_y^2 - \omega_E^2\right) \left(\omega_z^2 - \omega_E^2\right) \cos \phi_y   \nonumber \\ \nonumber \\
&-& E_{0x} \Omega_z \omega_E \left(\omega_y^2 - \omega_E^2\right) \left(\omega_z^2 - \omega_E^2\right) \sin \phi_x   \nonumber \\ \nonumber \\
&-& E_{0y} \gamma \omega_E \left(\omega_z^2 - \omega_E^2\right) \left(\omega_x^2 - \omega_E^2\right) \sin \phi_y   \nonumber \\ \nonumber \\
&+& E_{0z} \Omega_x \omega_E \left(\omega_x^2 - \omega_E^2\right) \left(\omega_y^2 - \omega_E^2\right) \sin \phi_z\Big\}   \;\;\;,
\end{eqnarray}

\begin{eqnarray}\label{anis25}
H_z &=& \frac{q}{m} \Big\{E_{0z} \left(\omega_x^2 - \omega_E^2\right) \left(\omega_y^2 - \omega_E^2\right) \left(\omega_z^2 - \omega_E^2\right) \cos \phi_z   \nonumber \\ \nonumber \\
&+& E_{0x} \Omega_y \omega_E \left(\omega_y^2 - \omega_E^2\right) \left(\omega_z^2 - \omega_E^2\right) \sin \phi_x   \nonumber \\ \nonumber \\
&-& E_{0y} \Omega_x \omega_E \left(\omega_z^2 - \omega_E^2\right) \left(\omega_x^2 - \omega_E^2\right) \sin \phi_y   \nonumber \\ \nonumber \\
&-& E_{0z} \gamma \omega_E \left(\omega_x^2 - \omega_E^2\right) \left(\omega_y^2 - \omega_E^2\right) \sin \phi_z\Big\}   \;\;\;,
\end{eqnarray}

\begin{eqnarray}\label{anis26}
H_x^\prime &=& \frac{q}{m} \Big\{E_{0x} \left(\omega_x^2 - \omega_E^2\right) \left(\omega_y^2 - \omega_E^2\right) \left(\omega_z^2 - \omega_E^2\right) \sin \phi_x   \nonumber \\ \nonumber \\
&+& E_{0x} \gamma \omega_E \left(\omega_y^2 - \omega_E^2\right) \left(\omega_z^2 - \omega_E^2\right) \cos \phi_x   \nonumber \\ \nonumber \\
&-& E_{0y} \Omega_z \omega_E \left(\omega_z^2 - \omega_E^2\right) \left(\omega_x^2 - \omega_E^2\right) \cos \phi_y   \nonumber \\ \nonumber \\
&+& E_{0z} \Omega_y \omega_E \left(\omega_x^2 - \omega_E^2\right) \left(\omega_y^2 - \omega_E^2\right) \cos \phi_z\Big\}   \;\;\;,
\end{eqnarray}

\begin{eqnarray}\label{anis27}
H_y^\prime &=& \frac{q}{m} \Big\{E_{0y} \left(\omega_x^2 - \omega_E^2\right) \left(\omega_y^2 - \omega_E^2\right) \left(\omega_z^2 - \omega_E^2\right) \sin \phi_y   \nonumber \\ \nonumber \\
&+& E_{0x} \Omega_z \omega_E \left(\omega_y^2 - \omega_E^2\right) \left(\omega_z^2 - \omega_E^2\right) \cos \phi_x   \nonumber \\ \nonumber \\
&+& E_{0y} \gamma \omega_E \left(\omega_z^2 - \omega_E^2\right) \left(\omega_x^2 - \omega_E^2\right) \cos \phi_y   \nonumber \\ \nonumber \\
&-& E_{0z} \Omega_x \omega_E \left(\omega_x^2 - \omega_E^2\right) \left(\omega_y^2 - \omega_E^2\right) \cos \phi_z\Big\}   \;\;\;,
\end{eqnarray}

\begin{eqnarray}\label{anis28}
H_z^\prime &=& \frac{q}{m} \Big\{E_{0z} \left(\omega_x^2 - \omega_E^2\right) \left(\omega_y^2 - \omega_E^2\right) \left(\omega_z^2 - \omega_E^2\right) \sin \phi_z   \nonumber \\ \nonumber \\
&-& E_{0x} \Omega_y \omega_E \left(\omega_y^2 - \omega_E^2\right) \left(\omega_z^2 - \omega_E^2\right) \cos \phi_x   \nonumber \\ \nonumber \\
&+& E_{0y} \Omega_x \omega_E \left(\omega_z^2 - \omega_E^2\right) \left(\omega_x^2 - \omega_E^2\right) \cos \phi_y   \nonumber \\ \nonumber \\
&+& E_{0z} \gamma \omega_E \left(\omega_x^2 - \omega_E^2\right) \left(\omega_y^2 - \omega_E^2\right) \cos \phi_z\Big\}   \;\;\;,
\end{eqnarray}

\begin{equation}\label{anis29}
H_1 = \Big\{\gamma \Omega_z \left(\omega_y^2 - \omega_E^2\right) \left(\omega_z^2 - \omega_E^2\right) + \gamma \Omega_z \left(\omega_z^2 - \omega_E^2\right) \left(\omega_x^2 - \omega_E^2\right) - \Omega_x \Omega_y \left(\omega_x^2 - \omega_E^2\right) \left(\omega_y^2 - \omega_E^2\right)\Big\} \omega_E^2   \;\;\;,
\end{equation}

\begin{equation}\label{anis30}
H_2 = \Big\{\gamma \Omega_y \left(\omega_x^2 - \omega_E^2\right) \left(\omega_y^2 - \omega_E^2\right) + \gamma \Omega_y \left(\omega_y^2 - \omega_E^2\right) \left(\omega_z^2 - \omega_E^2\right) + \Omega_z \Omega_x \left(\omega_z^2 - \omega_E^2\right) \left(\omega_x^2 - \omega_E^2\right)\Big\} \omega_E^2   \;\;\;,
\end{equation}

\begin{equation}\label{anis31}
H_3 = \Big\{\gamma \Omega_x \left(\omega_x^2 - \omega_E^2\right) \left(\omega_y^2 - \omega_E^2\right) + \gamma \Omega_x \left(\omega_z^2 - \omega_E^2\right) \left(\omega_x^2 - \omega_E^2\right) - \Omega_y \Omega_z \left(\omega_y^2 - \omega_E^2\right) \left(\omega_z^2 - \omega_E^2\right)\Big\} \omega_E^2   \;\;\;,
\end{equation}

\begin{equation}\label{anis32}
H_4 = \Big\{\gamma \Omega_z \left(\omega_y^2 - \omega_E^2\right) \left(\omega_z^2 - \omega_E^2\right) + \gamma \Omega_z \left(\omega_z^2 - \omega_E^2\right) \left(\omega_x^2 - \omega_E^2\right) + \Omega_x \Omega_y \left(\omega_x^2 - \omega_E^2\right) \left(\omega_y^2 - \omega_E^2\right)\Big\} \omega_E^2   \;\;\;,
\end{equation}

\begin{equation}\label{anis33}
H_5 = \Big\{\gamma \Omega_y \left(\omega_x^2 - \omega_E^2\right) \left(\omega_y^2 - \omega_E^2\right) + \gamma \Omega_y \left(\omega_y^2 - \omega_E^2\right) \left(\omega_z^2 - \omega_E^2\right) - \Omega_z \Omega_x \left(\omega_z^2 - \omega_E^2\right) \left(\omega_x^2 - \omega_E^2\right)\Big\} \omega_E^2   \;\;\;,
\end{equation}

\noindent
and

\begin{equation}\label{anis34}
H_6 = \Big\{\gamma \Omega_x \left(\omega_x^2 - \omega_E^2\right) \left(\omega_y^2 - \omega_E^2\right) + \gamma \Omega_x \left(\omega_z^2 - \omega_E^2\right) \left(\omega_x^2 - \omega_E^2\right) + \Omega_y \Omega_z \left(\omega_y^2 - \omega_E^2\right) \left(\omega_z^2 - \omega_E^2\right)\Big\} \omega_E^2   \;\;\;.
\end{equation}

\end{document}